\documentclass[aps,prl,amsmath,amssymb,showpacs,twocolumn]{revtex4}
\usepackage{bbold}
\usepackage{graphicx}
\usepackage{amsmath}
\usepackage{amssymb,amsthm}

\usepackage{hyperref}
\usepackage{color}

\def\eq{\begin{eqnarray}}
\def\en{\end{eqnarray}}

\usepackage[normalem]{ulem} 
\usepackage{ragged2e}

\usepackage[caption=false,position=t,singlelinecheck=false,justification=raggedright]{subfig}
\usepackage{enumitem}

\begin{document}
\title{Virtual learning environment for interactive engagement with advanced quantum mechanics
}
\author{Mads~Kock~Pedersen$^1$, Birk~Skyum$^1$, Robert~Heck$^1$, Romain~M\"{u}ller$^1$, Mark~Bason$^2$, Andreas~Lieberoth$^1$ and Jacob~F.~Sherson$^1$}
\email{sherson@phys.au.dk}
\affiliation{$^1$ AU Ideas Center for Community Driven Research, CODER, \\
 Department of Physics and Astronomy, Aarhus University, DK--8000 Aarhus C, Denmark \\
 $^2$ School of Physics and Astronomy, University of Nottingham, University Park, Nottingham NG7 2RD, United Kingdom}

\date{\today}

%%%%%%%%%%%% Abstract %%%%%%%%%%%%%%

\begin{abstract}
A virtual learning environment can engage university students in the learning process in ways that the traditional lectures and lab formats can not. We present our virtual learning environment \emph{StudentResearcher} which incorporates simulations, multiple-choice quizzes, video lectures and gamification into a learning path for quantum mechanics at the advanced university level. \emph{StudentResearcher} is built upon the experiences gathered from workshops with the citizen science game Quantum Moves at the high-school and university level, where the games were used extensively to illustrate the basic concepts of quantum mechanics. The first test of this new virtual learning environment was a 2014 course in advanced quantum mechanics at Aarhus University with 47 enrolled students. We found increased learning for the students who were more active on the platform independent of their previous performances.
\end{abstract}

%----------------------- end Abstract---------------------
\maketitle

\section*{Introduction}
Traditionally, teaching in physics at the university level is dominated by lectures and lab exercises. However, lectures are limited in their effectiveness of conveying certain kinds of knowledge, since students are passive participants \cite{Mazur1997}. It has been shown that a learning environment in which students are active participants can more efficiently develop students' competences and increase their information retention \cite{Mazur1997,Wieman2005}. To transform physics classes into an active learning environment, we can change the format of the lectures \cite{Wieman2011,Schwartz1998}, and we can offer the students new virtual learning environments and methods to enhance their studies \cite{Mayer1998}.

Virtual learning environments (VLE) can have different design philosophies. For instance, the \emph{Institute of Physics New Quantum Curriculum} \cite{Kohnle2014} teaches quantum mechanics through a series of texts and simulations. It was built on established \emph{PhET Look and Feel} design principles \cite{Adams2008a,Adams2008b}, which encourage the use of an open and exploratory design for simulations. Other VLE's also use simulations to teach quantum mechanics in a similar manner \cite{Gawron2015,Bronne2009,Garcia2000,Lawrence1996,Zollman2002}.
Another direction is \emph{Peerwise} \cite{Denny2008}, which allows students to author and answer each other's multiple-choice questions in a peer-instruction format. \emph{Peerwise} forces the students to come up with plausible wrong answers. Thus, students need to both know the curriculum well and to reason about what other students would find tricky and/or important.

We want to build a system which uses gamification, i.e., the use of game-like elements \cite{Groh2012}, to convey the curriculum in ways different from the dominant lecture and lab based formats \cite{Bonde2014}. Being given the opportunity to solve problems in a gamelike environment with wide berth for trial and error provides a sense of the relationships between interacting elements, which can otherwise be difficult to express in traditional didactic formats \cite{Biesta2010}. For such learning to be effective, however, the interactions have to be presented with purpose \cite{Bennett1996}, and students allowed to reflect on the resulting experiences in conjunction with their overall learning trajectories \cite{Dewey1938}.

\begin{figure}[t]
\includegraphics[width=0.4\columnwidth]{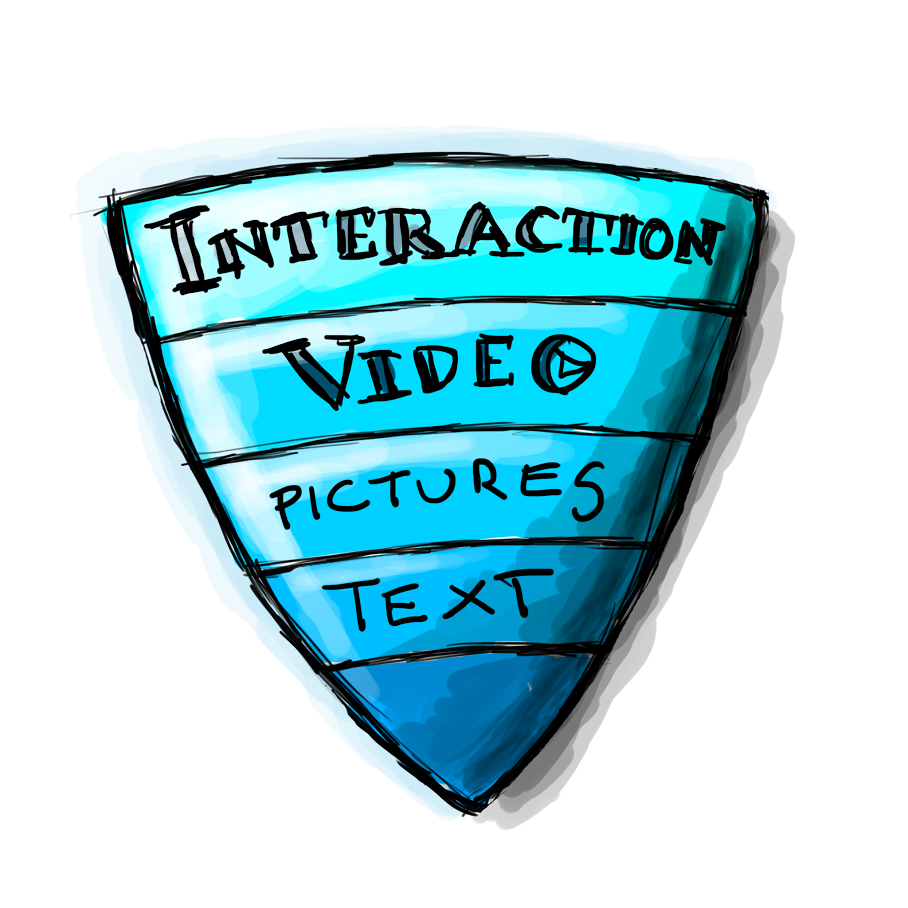}
\includegraphics[width=0.55\columnwidth]{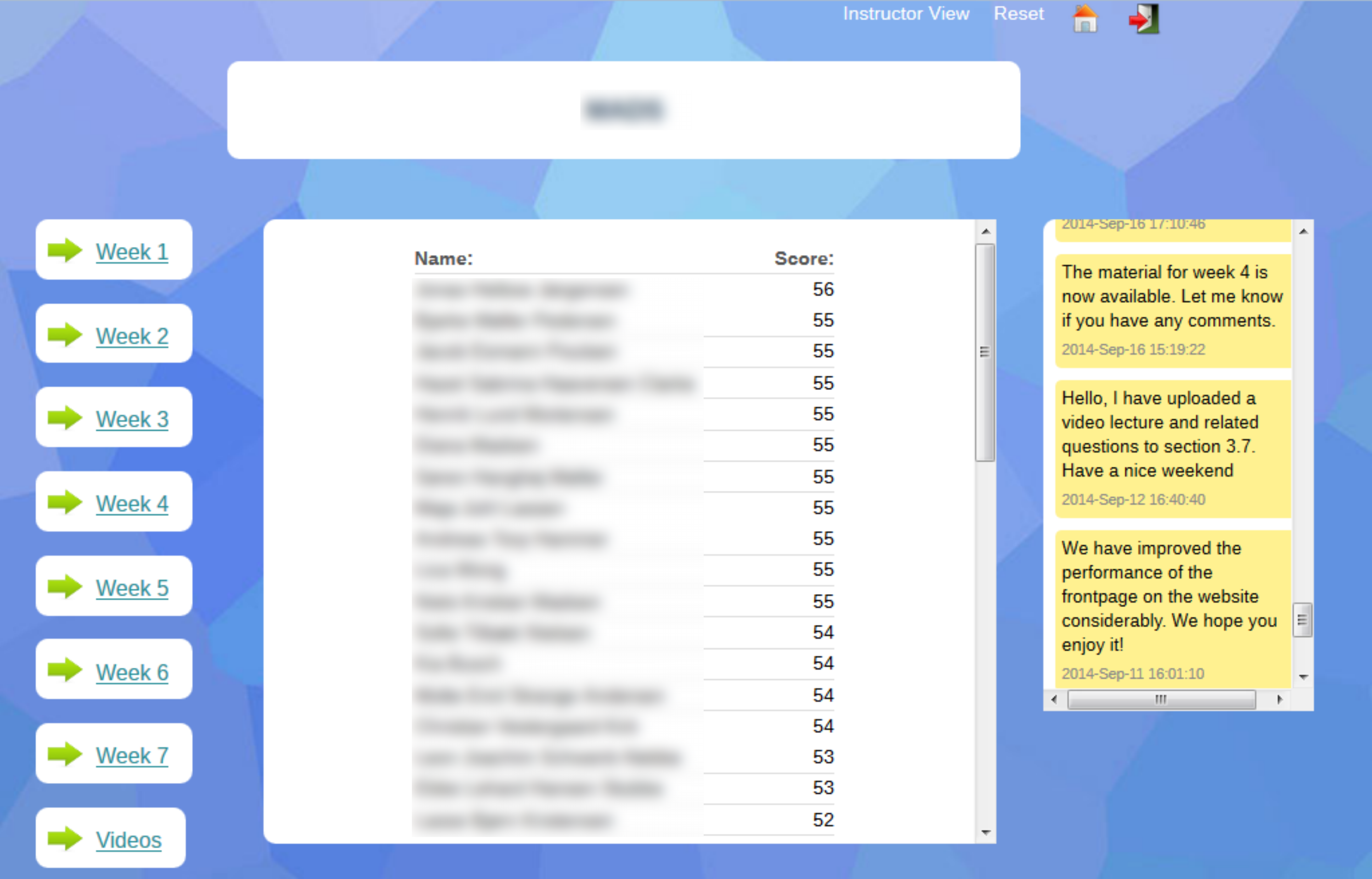}
\centering
\caption{(Color online) Left: Content priority for \emph{StudentResearcher}. Convey most of the content through interactive elements, then video format, illustrations or text in descending order. Right: The landing page of \emph{StudentResearcher} after login.}
\label{fig1}
\end{figure}

Our team conducted a series of game-based workshops at high-school \cite{Magnussen2014,Lieberoth2015} and university \cite{Bjaelde2014} levels. The citizen science game \emph{Quantum Moves} \cite{Lieberoth2014,Soerensen2015} with the underlying challenge of building a quantum computer \cite{Weitenberg2011} was used in the workshops as a contextual and motivational background for teaching Newtonian and basic quantum mechanics.  The call for aid in the citizen science games was instrumental in shifting the students from passive consumers of static knowledge to active co-creators of new knowledge \cite{Lieberoth2015}. Building on this we want to establish a scalable structure, which can convey the combined elements used in these workshops. Thus, we built \emph{StudentResearcher}\cite{Web}, which includes a mixture of interactive elements, such as brief games, video lectures, and graphical information [Fig. \ref{fig1}].

Insights from our early workshops were turned into guiding principles for \emph{StudentResearcher}. In terms of pedagogical design, this was an opportunity to conduct an exploration of student trajectories with autonomous access to a supplementary diet of interactive problems and 3D simulations unfettered by any particular pedagogical hypotheses. The first test of \emph{StudentResearcher} was its deployment as an extensive supplement to traditional lectures and theoretical exercises in a 7-week graduate-level advanced quantum mechanics course at Aarhus University from August to October 2014. All 47 participants enrolled were encouraged, but not forced, to actively use \emph{StudentResearcher} as part of their learning trajectory. Beyond an immediate introduction to the new tools, little was changed from how the course was conducted in the previous two years. At the end of the course the students had to take an oral exam. Thus, the content was designed to make the student actively reflect upon the curriculum, and to further encourage the students to reflect on the curriculum we also integrated \emph{Peerwise} \cite{Denny2008} into our VLE.

We present the design principles of \emph{StudentResearcher's} content. Based on data gathered from our case study, we analyse the extent, to which \emph{StudentResearcher} was used by the students, as well as how it predicted exam performance. The data is compared to that gathered from the same course in 2013 which had the same lectures and exercises, but did not feature \emph{StudentResearcher}. We will answer the following research questions:
\begin{enumerate}[label={\text{RQ.}\theenumi}]
\item How did students perceive the interactive elements of \emph{StudentResearcher}?
\item Did students exposed to \emph{StudentResearcher} in 2014 perform better in the midterm test than the students in 2013?
\item Did students who were more active on \emph{StudentResearcher} perform better at the exam?
\item Can data from \emph{StudentResearcher} be used to identify a disconnect between the core curriculum and the students' abilities?
\end{enumerate}

\section*{Content and design principles in \emph{StudentResearcher}}

\begin{figure*}[t]
\includegraphics[trim=80 0 80 0, clip=true, height=0.3\textwidth]{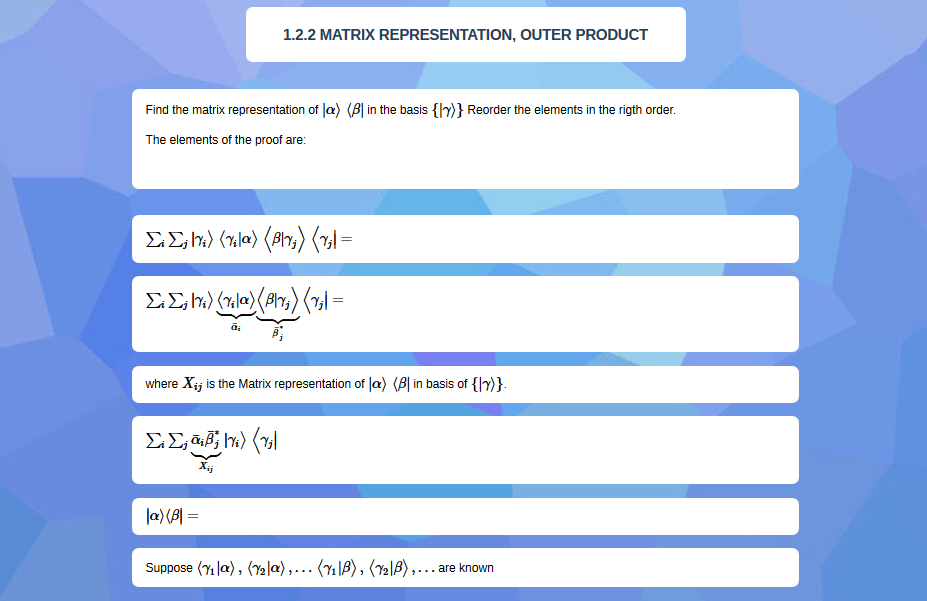} \qquad
\includegraphics[height=0.3\textwidth]{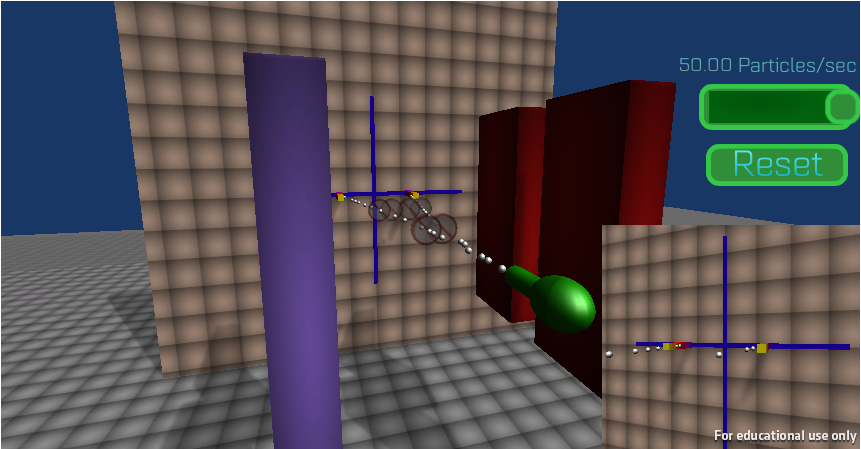}
\caption{(Color online) Left: Example of a reshuffling exercise. The students had to drag the steps of the proof such that they were in the correct sequence. Right: Screenshot of the SGE simulator. The simulator allows the student to change the orientation of the magnets, to change the rate of particles, and to block certain spin components. When the particles hit the back-screen, a histogram of where the particles hit is generated.}
\label{fig2}
\end{figure*}

Our a-priori goal was to create an active and fluid environment, wherein learners could engage with the curriculum on multiple cognitive dimensions. We implemented a points-badges-leaderboard framework, which awarded points to the users for completing tasks, and a weekly badge for exceeding a threshold number of points. The students were aware that neither the points nor the badges would have direct influence on their grades, but just as the conventional theoretical exercises \emph{StudentResearcher} was framed as a part of the course material in preparation for the oral exam.

A leaderboard displaying high scores of the interactive elements was placed at the landing page [Fig. \ref{fig1}]. The students could opt-out of appearing on the leaderboards seen by classmates to avoid deterring anyone from participating in \emph{StudentResearcher} due to feelings of negative exposure. To ensure a sense of seriousness the students' real names were used.

\emph{StudentResearcher} content was divided into weeks in accordance with scheduled lectures. Each week consisted of multiple voluntary \emph{StudentResearcher} \emph{lessons}. Each such lesson was a manageable chunk of the curriculum requiring 10-20 minutes to complete. This personal autonomy allowed students to move through the material at their own pace. At the end of each week the students were asked to fill in text fields describing the hardest and the most important parts of the curriculum that week. This allowed students to reflect on the curriculum, and convey a sense of co-creation. 

A badge was awarded when a student exceeded a set number of points during a week. This threshold fell between 50-67\% of the total points available each week. 

In the following we will present the main features of the most crucial elements of \emph{StudentResearcher} beyond multiple-choice tests.

\emph{Reshuffling proofs --}
Since the course exam was oral, the ability to understand and explicitly express the logical and domain-specific reasoning at each step of a proof was essential to performance. To train these abilities we created an interactive module which provided all the steps of a proof in random order [Fig. \ref{fig2}]. Students then had to give the correct sequence of the derivation. The reshuffling exercises enabled a re-emphasis of important derivations and were an alternative to going through all proofs at the lectures.

\emph{Simulations --}
Simulations and visualizations of quantum physics experiments and concepts have proven to be effective in conveying the curriculum \cite{Kontogeorgiou2008,Bobroff2013,Schroeder1993}. Thus, we built a 3D simulation of the Stern-Gerlach experiment (SGE), which is central to the course, because it illustrates the counter-intuitive quantum effects appearing when a spin is measured along orthogonal axes \cite{Sakurai2011}. The simulation was built in the \emph{Unity} development environment and hosted in a specific lesson in \emph{StudentResearcher}. The simulation displayed a stream of particles in a fifty-fifty mixture of the spin-up and spin-down states going through up to three sets of rotatable magnets  [Fig. \ref{fig2}]. Each set of magnets altered the path of each particle based on the spin. If a screen was placed after a set of magnets, the spin of the particles could be determined. Simulations of the SGE enable students to apply the theory in a spatial setting, and highlights the statistical nature of quantum physics \cite{Schroeder1993,Guangtian2011}.

The first level of the 3D simulation contained only one set of magnets. Students were able to control whether the magnets were horizontally or vertically oriented. The simulation continued through the lesson to add more sets of magnets and allowed students to choose to block specific spin components after each set of magnets. The challenges in the simulations were either to predict the pattern on the back-screen given a specific configuration of magnets and blockers, or to generate a specific pattern on the back-screen by creating a specific configuration. This allowed students to gain an intuition for the system in a puzzle-like activity.

The quantum mechanics simulation tool was used in a simplified form in high-schools  \cite{Magnussen2014,Lieberoth2015}. The full version was introduced in a 2$^\text{nd}$ year university introductory quantum mechanics course \cite{Bjaelde2014}, as well as in this 4$^\text{th}$ year course with more emphasis on time dependent dynamics. It allowed students to experiment with specifying potentials and examining eigenstate mixtures in the context of the time independent and dependent Schr\"{o}dinger equation. The time evolution of the designed wave-function could be performed in an auxiliary potential. This helps students visualize that the time evolution in a static potential can be explained by a phase evolution of the eigenstates. In addition, it presented an opportunity to bring together otherwise unrelated features of the core curriculum (parity selection rules and time dependent perturbation theory) to achieve a detailed understanding of the core research challenge of the \emph{Quantum Moves} games -- to remove kinetic energy from an oscillating cloud.

\emph{Peerwise --}
In the sixth week of the course we asked the students to use \emph{Peerwise} \cite{Denny2008}. \emph{Peerwise} was presented during a lecture, where examples of good and bad \emph{Peerwise} questions were given. After practicing at the lecture, students were given the task of authoring their own questions at home and awarded points for both authoring and answering questions. \emph{Peerwise} has its own built in scoring, rating, and ranking system. However, we manually imported the points into \emph{StudentResearcher} to give the students the experience of a seamless integration of all activities.

\begin{figure*}
\includegraphics[trim=25 40 10 35 ,clip=true, height=0.3\textwidth]{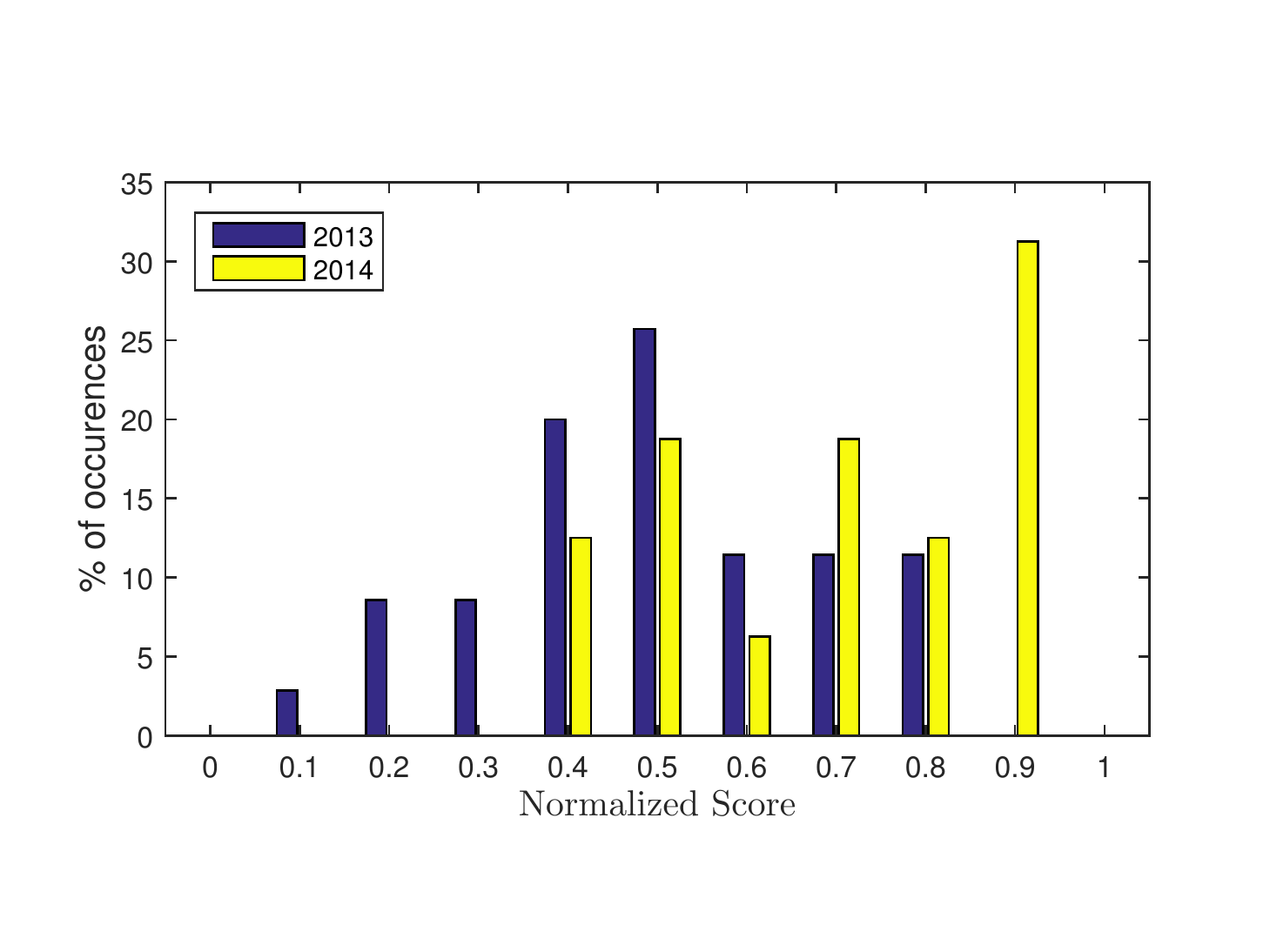}
\includegraphics[trim=-30 10 30 55 ,clip=true, height=0.27\textwidth]{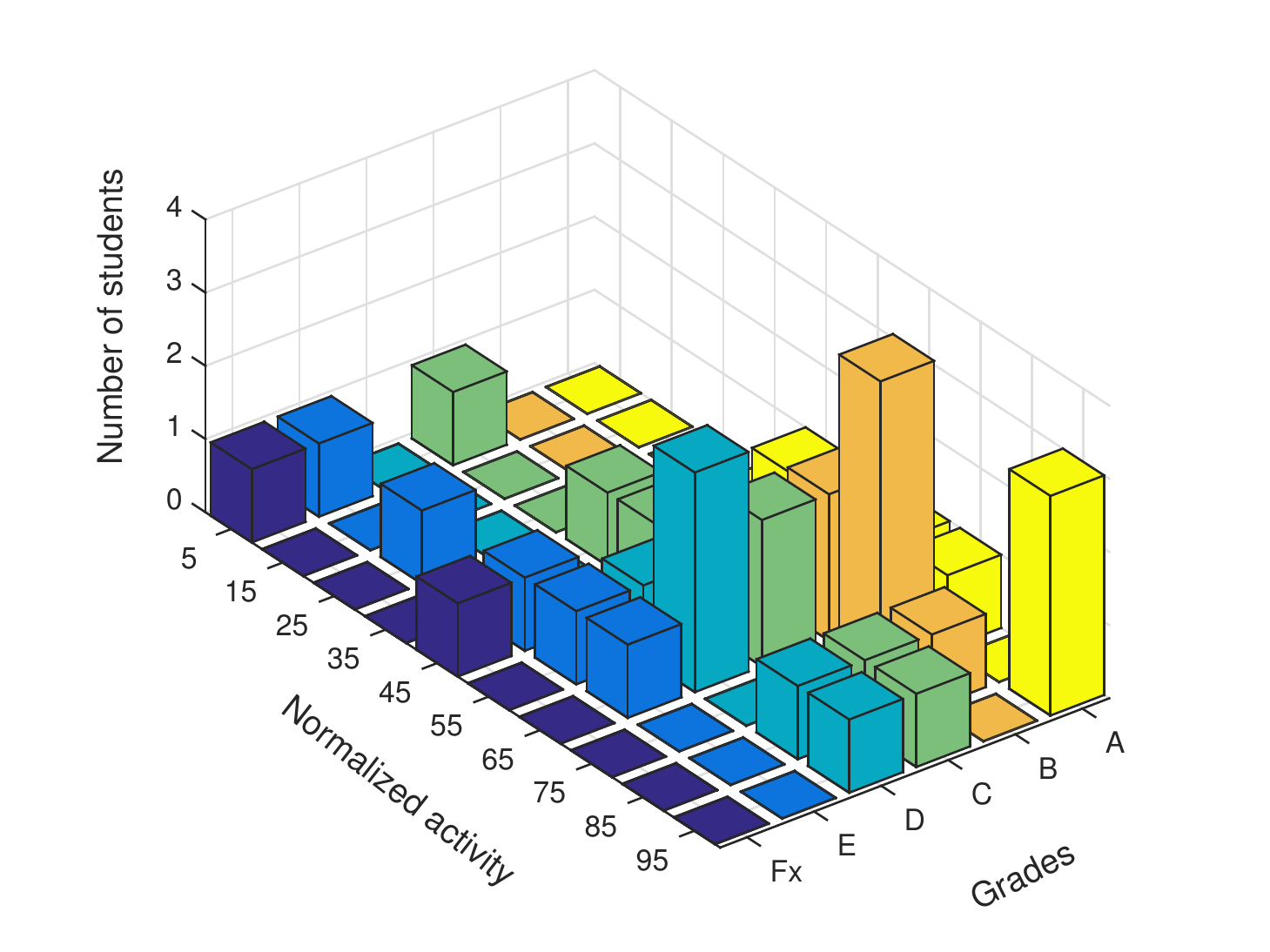}
\centering
\caption[caption]{(Color online) Left: Results from the midterm test in both 2013 (blue) and 2014 (yellow). In 2013 there was 35 students who answered the midterm test, whereas 17 of the 47 students \cite{Endnote1} answered the midterm test in 2014.  Right: Histogram of students' activity on \emph{StudentResearcher} vs. the grade they received at the oral exam.}
\label{fig3}
\end{figure*}

\emph{StudentResearcher} consisted of 192 slides, 7 video lectures, 4 SGE games, 1  quantum physics simulation tool, 24 reshuffling exercises, 162  multiple-choice questions, and 25 questions authored by students in \emph{Peerwise}. Examples of the different forms of content can be found in the Supplementary Appendix. In the final week the students rated the education value of the different interactive elements on a 1-5 Likert scale.

An optional and anonymous midterm test was administered in both 2013 and 2014 in the fifth week of the courses. The written test consisted of 10 questions including definitions, small derivations, and calculations.  Each answer was graded by the instructor on a scale from 0 to 3 points: 0 completely incorrect, 1 mostly incorrect, 2 mostly correct, 3 completely incorrect.

At the oral exam students were questioned on the textbook materials, the theoretical and VLE exercises \cite{CourceDescription}.  

\section*{Findings}
Of the 47 students initially enrolled, 12 opted-out of the leaderboard. There were no distinguishable differences between the users who opted-out and those who stayed.

\begin{figure}[b!]
\includegraphics[trim=25 0 40 0 ,clip=true, width=1\columnwidth]{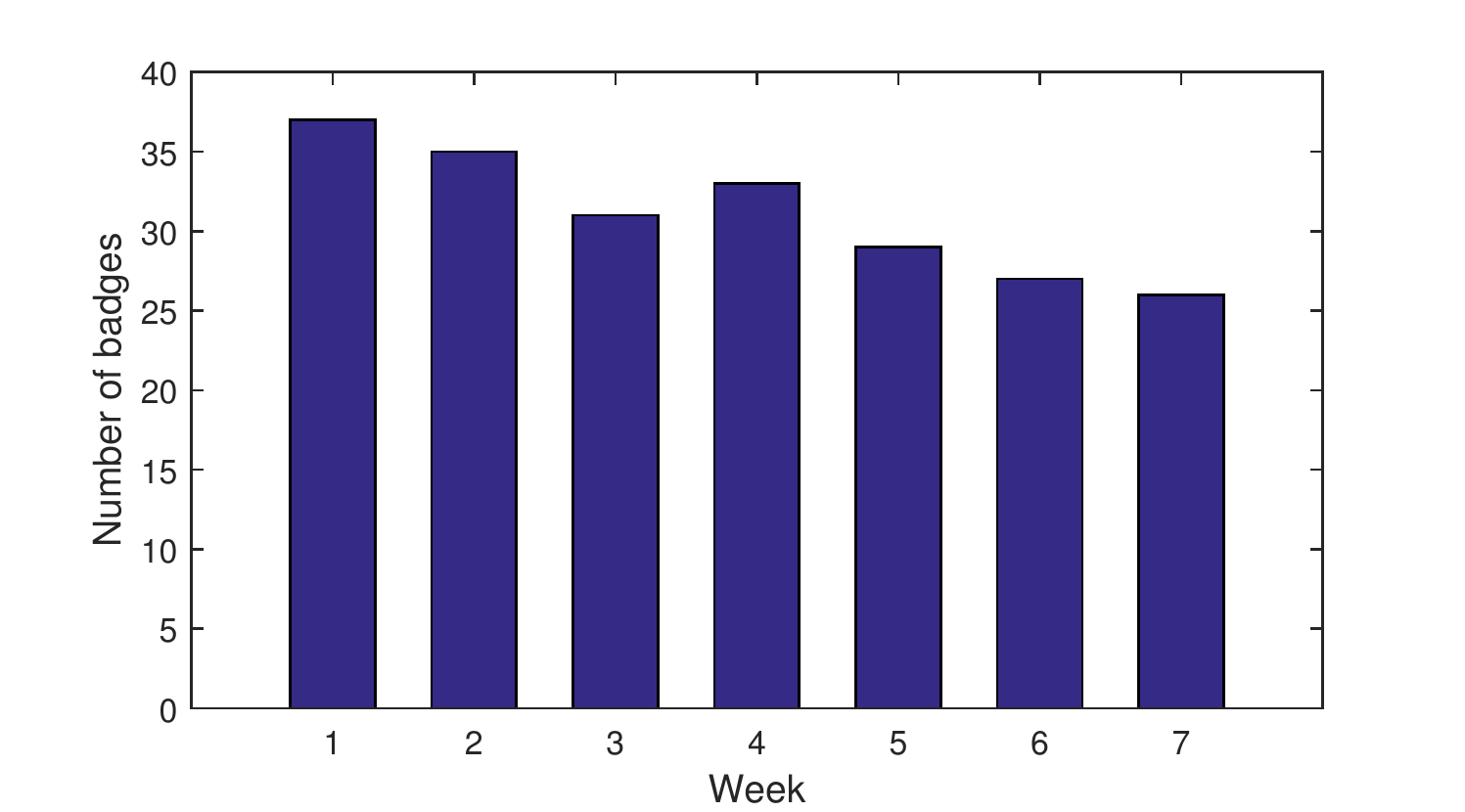}
\centering
\caption{(Color online) Number of badges awarded each week during the course. Each user could only receive one badge per week.}
\label{fig4}
\end{figure}

In order to test how students perceived the interactive elements [RQ.1], we looked at the students' evaluations, and compared the 5 interactive element types. A Kruskal-Wallis test found a significant between-distributions difference. Thus, a Mann-Whitney U post-hoc test was applied, to determine which distributions differed from each other. This revealed that students found the multiple-choice tests (Mean=4.4) significantly more rewarding than the simulations ($M=3.4$, $z=-3.18$ $p=0.001$) with a high-medium effect size ($r=0.47$) \cite{Cohen1988}. Overall, the students felt that \emph{StudentResearcher} was rewarding (Mean=4.1).

To test the effect of introducing \emph{StudentResearcher} [RQ.2], we compared the midterm-test results from the class of 2014 who were exposed to \emph{StudentResearcher} with the class of 2013 who were \emph{not} exposed to \emph{StudentResearcher}.  A Mann-Whitney U test revealed that students in the 2014 cohort  performed significantly better in the midterm test than their peers from 2013 ($z= -3.30 $, $p < 0.001$) with a high-medium effect size ($r=0.45$) [Fig. \ref{fig3} Left]. 

In order to test whether students who actively used  \emph{StudentResearcher} performed better at the exam [RQ.3], we computed the correlation ($\rho$) between grades \cite{Endnote2} and the total number of times students had used the interactive elements [Fig. \ref{fig3} Right], i.e., their activity. This revealed a strong correlation ($\rho=0.55$, $p = 0.002$) between activity on \emph{StudentResearcher} and the grade received at the oral exam. As expected there was already a strong correlation between course grade and the students’ overall GPA ($\rho=0.62$, $p<0.001$), but GPA and \emph{StudentResearcher}-activity were not statistically correlated. The correlation between course grade and \emph{StudentResearcher}-activity even remained significant when controlling for effects of GPA ($\rho=0.45$, $p=0.03$, $N=24$).

During the course we used the \emph{StudentResearcher} data to identify a disconnect between students' actual abilities to perform change of basis calculations and the expected level of competence [RQ.4]. We collected data on 11 online questions about change of basis, and found that the average error percentage was 79\% ($SD=7$)! This was then addressed during lectures, and for the three questions  about change of basis in the midterm-test with a max score of 9 students from 2014 ($M = 7.13$) significantly outperformed students from 2013 ($M = 4.60$, $z=-2.61$ $p=0.009$), with a medium effect size ($r=0.37$).

\section*{Discussion}

The main difference between the 2013 and 2014 courses was \emph{StudentResearcher}. We found that students in 2014 who were exposed to \emph{StudentResearcher} significantly outperformed the 2013 students. Thus, we interpret differences in test results as a direct effect of the new interactive learning opportunities. However, the difference could also stem from other sources such as classroom factors or cohort composition. Likewise, the effects from \emph{StudentResearcher} on the grade received may be moderated by general background variables such as overall diligence and study skills, but we still established  a clear statistical link between activity on \emph{StudentResearcher} and course grade. Together with the finding that activity on \emph{StudentResearcher} was not correlated with the GPA, we conclude that the students' use of \emph{StudentResearcher} improves their learning, and that the effect can be seen for both weaker and stronger students.

One particular advantage of \emph{StudentResearcher} is the instantaneous feedback offered upon answering. This helps students identify their own cognitive disconnects between topics that superficially seem simple, but hide subtleties, when explicit calculations have to be made. A prime example from this course is the transformation of two-level superposition states from one basis to another. In previous years this had been treated abstractly at lectures and very briefly at joint theoretical exercises. The integration of practical calculation in \emph{StudentResearcher} revealed a remarkably widespread conceptual disconnect in transferring from abstract knowledge to concrete calculations. Thus, \emph{StudentResearcher} exercises helped the lecturer conceptualize this disconnect. At lectures, this knowledge was used to support much deeper coverage of the subject in the following weeks. Compared to 2013 this, combined with the resulting increase in focus on the topic, constitutes a reasonable explanation for the statistically significant increase, we were able to detect, in proportion of correct answers in the test on this topic.

Since using \emph{StudentResearcher} was voluntary a certain drop-off in activity was expected as the exam approached. The number of badges achieved [Fig. \ref{fig4}] can be taken as an indication of how diligently students used \emph{StudentResearcher}. This number was only reduced by 30\% in the last week compared to week one, which is very good compared to the retention in citizen science games on the internet \cite{Lieberoth2014}.

 Survey research revealed that students in workshops using citizen science games found the exercises fun, but did not feel they had learned much, even though a pre-post test showed that they had actually improved \cite{Magnussen2014, Bjaelde2014}. We found the same pattern here with student preferring multiple-choice tests above the simulations. This can stem from students feeling that they fail to gain understanding from the simulations opposed to multiple-choice tests, which more closely match the explicit format used in traditional teaching practices and the oral exam procedures. Thus, any implicit knowledge obtained from the simulations may be opaque to students. It is very likely that this blindness is a cultural product of our traditional semantically explicit \cite{Conway1997} and highly exam-focused teaching traditions \cite{Biesta2010,Schank2011}. Even though the simulations were not rated as highly as the practical activities, we observed that these activities gave rise to many more discussions on the foundations of quantum mechanics, such as interpretations of SGE, than had taken place the previous year. This represents great educational value, since the underlying purpose of the course was to give an axiomatic presentation of quantum mechanics and to spark a discussion of the chosen axioms. Although anecdotal, these observations hint at the value of game-based exercises for facilitating discussions on the more implicit knowledge hidden in any curriculum. 

\section*{Conclusions and outlook}
\emph{StudentResearcher} was well received by our students, and midterm test scores improved from 2013 to 2014 when \emph{StudentResearcher} was introduced. We found a notable correlation  between students' activity in \emph{StudentResearcher} and their exam grade, even when controlling for their overall GPA's. This demonstrates the value of \emph{StudentResearcher} as a supplement to traditional lectures and homework when dealing with elements in advanced quantum physics which can be hard to represent in non-interactive learning formats.

Future studies toward individualized learning could investigate students' different motivations with respect to activity and learning outcomes.

\emph{Acknowledgements --} The authors would like to thank Paul Denny for assistance in the integration of the \emph{Peerwise} system into \emph{StudentResearcher} and Pernille Maj Svendsen and Ole Bj\ae lde for helpful comments. Financial support from the Aarhus University Research Foundation and the John Templeton Foundation is gratefully acknowledged.

\bibliographystyle{h-physrev}

\newpage
\begin{minipage}{\textwidth}
\begin{center}
\section*{Supplementary Appendix}
\end{center}
\end{minipage}
\\
\begin{minipage}{\textwidth}
\emph{Multiple choice questions}
\begin{center}
$\vcenter{\hbox{\includegraphics[width=0.40\textwidth]{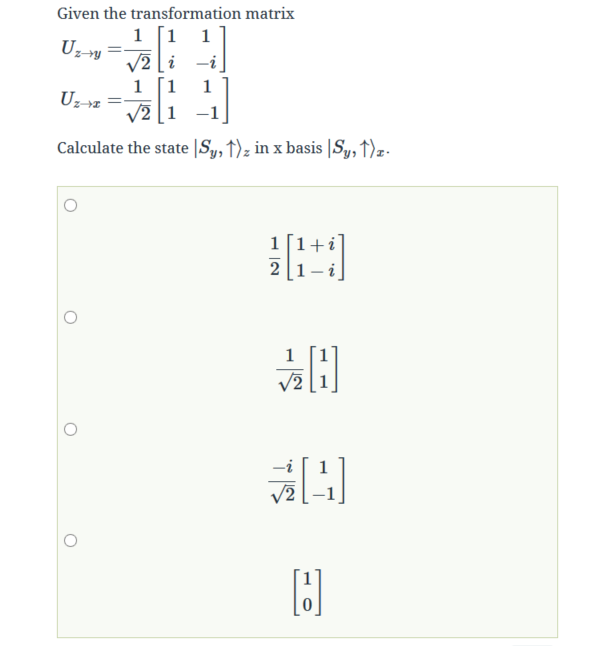}}}$
$\vcenter{\hbox{\includegraphics[width=0.40\textwidth]{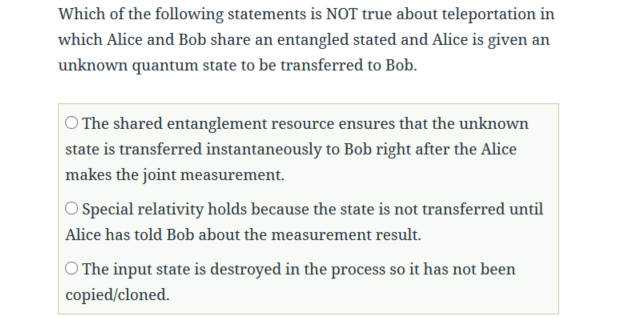}}}$

\vspace{0.05\textwidth}

$\vcenter{\hbox{\includegraphics[width=0.4\textwidth]{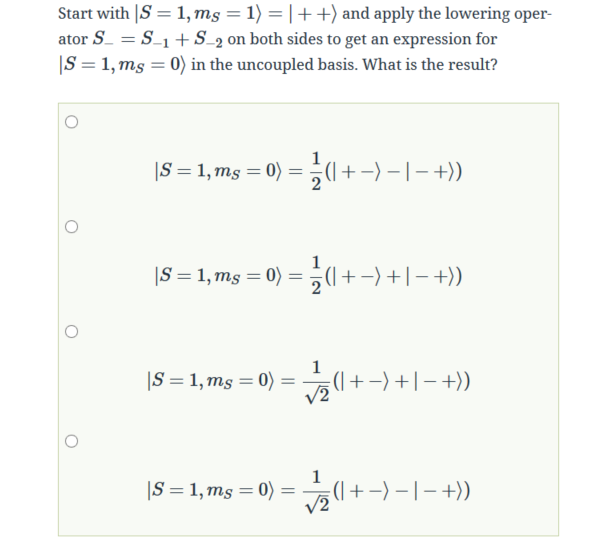}}}$
$\vcenter{\hbox{\includegraphics[width=0.4\textwidth]{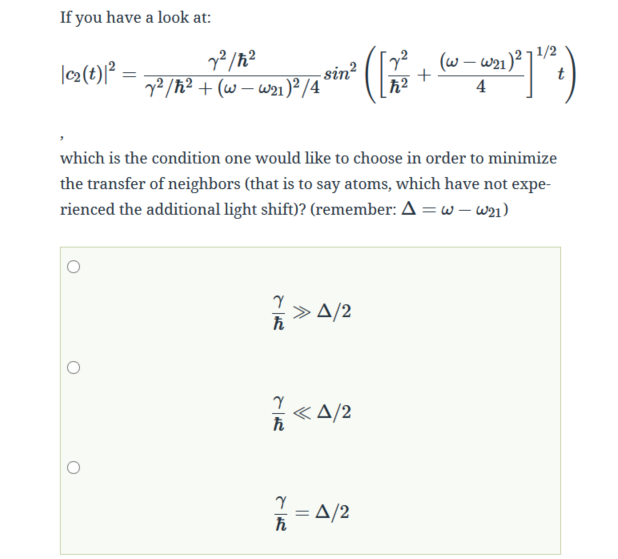}}}$
\end{center}
FIG. 5: (Color online) Four examples of multiple-choice questions from \emph{StudentResearcher}. Top-Left:State transformations. Top-Right: Quantum teleportation. Bottom-Left: Clebsch-Gordan coefficients in a spin 1/2 system. Bottom-Right: Single atom manipulation.
\end{minipage}
\clearpage
\newpage
\begin{minipage}{\textwidth}
\emph{Reshuffling proofs}
\begin{center}
$\vcenter{\hbox{\includegraphics[width=0.4\textwidth]{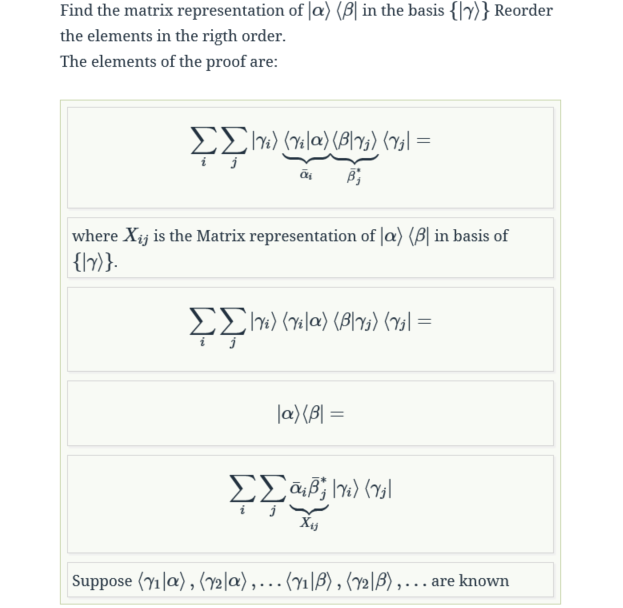}}}$
$\vcenter{\hbox{\includegraphics[width=0.4\textwidth]{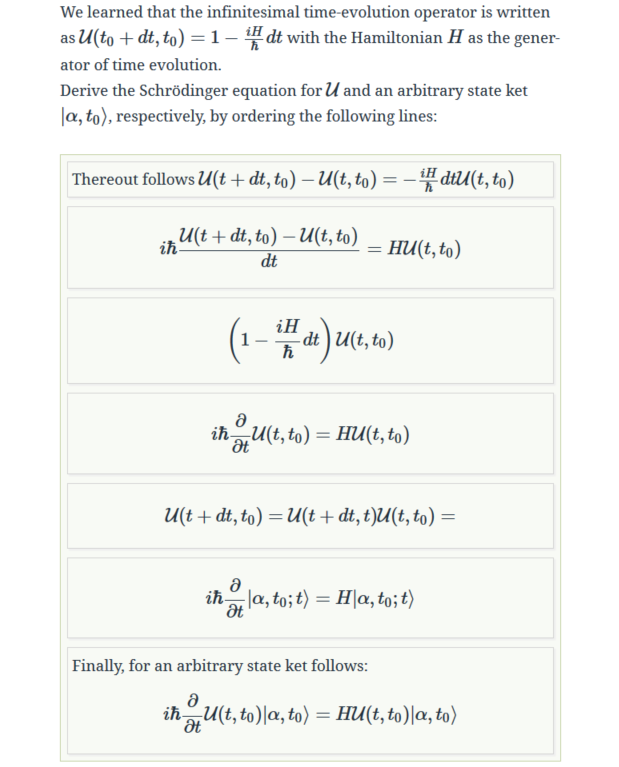}}}$

\vspace{0.02\textwidth}

$\vcenter{\hbox{\includegraphics[width=0.4\textwidth]{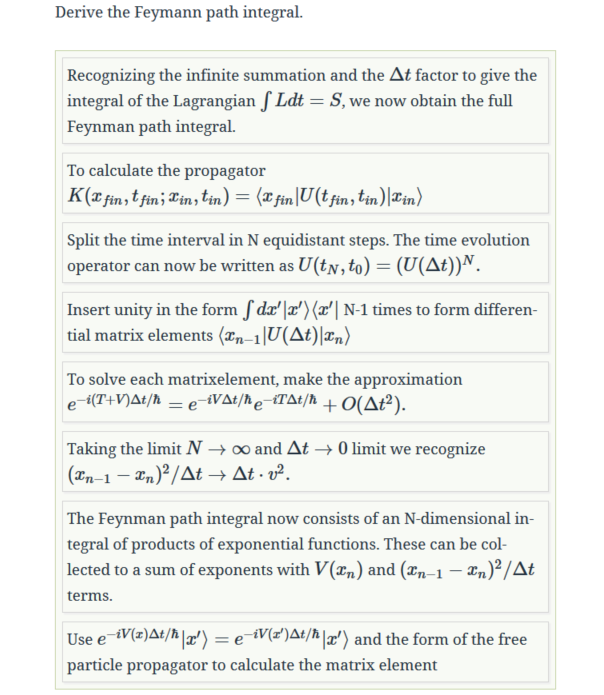}}}$
$\vcenter{\hbox{\includegraphics[width=0.4\textwidth]{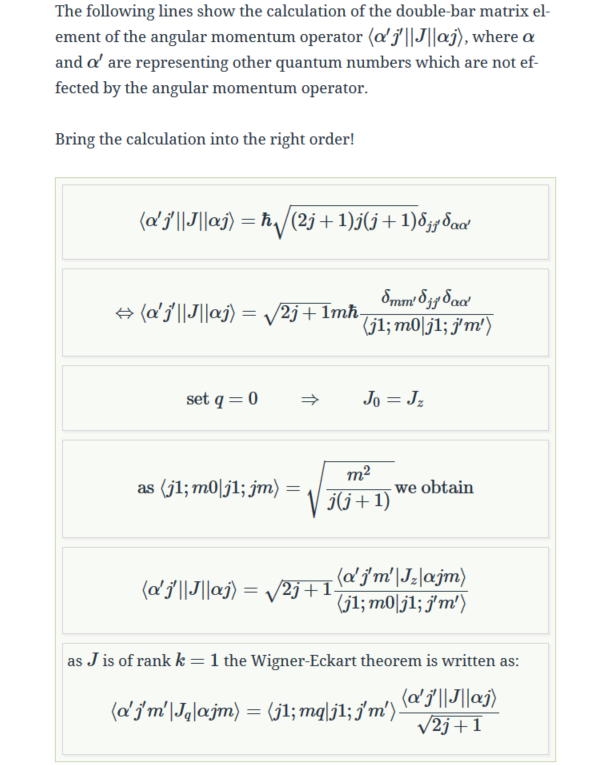}}}$
\end{center}
FIG. 6: (Color online) Four examples of reshuffling proof questions from \emph{StudentResearcher}. Top-Left:Matrix representation, Outer product. Top-Right: Derivation Of The Schrödinger Equation. Bottom-Left: Derivation of Feynman path integral. Bottom-Right: Applications of the Wigner-Eckart theorem.
\end{minipage}

\end{document}